# Bremermann's limit in *cGh*-physics

*Gennady Gorelik,* gennady.gorelik@gmail.com

Center for Philosophy and History of Science, Boston University, retired

## ABSTRACT

Do physical laws limit the speed of "*all data processing systems, manmade as well as biological*"? This question proposed and positively answered by H. J. Bremermann in 1962, should be corrected to make it compatible with Einstein's theory of gravity (aka General Relativity, or GR). As a result, the *Bremermann's limit*, proportional to mass *M* of any 'computer,' $Mc^2/h = \sim (M/\text{gram})10^{47}$ bits per second, should be replaced by the absolute limit $(c^5/Gh)^{1/2} = \sim 10^{43}$ bits per second, where the universal constants *c, G,* and *h* are the speed of light, the gravitational constant, and Planck's constant.

KEYWORDS: data processing rate, Bremermann's limit, general relativity, universal physical constants, Planck values, quantum gravity

Do physical laws limit the speed of computation, or bitrate, of '*all data processing systems, manmade as well as biological*' (hereafter referred to as 'computer')? This question was asked by H. J. Bremermann back in 1962, and his positive answer was the so-called Bremermann's limit $Mc^2/h = \sim (M/\text{gram})10^{47}$ bits per second, where *M* is the mass of the computer, *c* is the speed of light, and *h* is Planck's constant.[1] Bremermann derived this limit from two basic physical relations $E = Mc^2$ and $\Delta E \Delta t > h$ , but he didn't discuss why his formula contained only these two physical constants.

There are many physical constants that are *fundamental* (i.e., cannot be reduced to other constants by any existing theory) and must be measured experimentally. Only three fundamental constants can be called *universal*, since they must in principle be taken into account when describing any physical phenomenon. They can be neglected only if the accuracy of the quantitative description is sufficiently limited. These universal constants are *c, h* and the gravitational constant *G*. They participate in the basic equations of the fundamental theories of physics: relativity, quantum mechanics and Einstein's theory of gravity. This *cGh*-view of the history and interrelation of physical theories was developed in the 1930s by Matvei Bronstein (1906-38).[2]

Bremermann noted that the maximum bitrate of a computer with a mass equal to the mass of the Earth ($10^{75}$ bps) is small compared to the number of all sequences of move in chess or the patterns in a black and white mosaic of 100 × 100 cells. However, while bitrates $10^{47}$ and $10^{75}$ are not high enough for some computing tasks, the inverse values, i.e. the duration of one operation – $10^{-47}$ and $10^{-75}$ sec, are unreasonably small for theoretical physics, if we have in mind the theory of quantum gravity. Such a has not yet been created, but some characteristic values of physical parameters are well known, beyond which modern physical theories are inapplicable. These characteristic ***Planck values*** are constructed from three universal constants *c, G, h*, and, in particular, the Planck time $\Delta t_{cGh} = (hG/c^5)^{1/2} = 10^{-43}$ sec is well above the mentioned durations of one operation (which should be implemented in some physical process). [3]

To deal with this contradiction let's revisit Bremermann's way to his limit. He used the relativistic formula $E = Mc^2$, but ignored the size of computer and therefore implicitly assumed an infinite speed of signal propagation within the computer, which contradicts the theory of relativity.

To obtain the minimum time of one operation $\Delta t$ in a computer with mass $M$ and linear size $L$, we should combine quantum constraint $\Delta t > h/\Delta E$, relativistic constraint of available energy $\Delta E < Mc^2$ and require that the speed of signal within computer does not exceed the speed of light $L/\Delta t < c$ . Therefore, $\Delta t > \max [h/Mc^2, L/c]$. This would lead to Bremermann's limit $\Delta t_B = h / Mc^2$, if $L$ could be chosen as small as desired: $L < h/Mc$.

However, one is not free in choosing the values of $L$ and $M$, and here the third universal - gravitational - constant $G$ appears. To prevent the computer from turning into a "black hole" and disappearing beyond the event horizon, the condition $L > GM/c^2$ must be met. The combination of these restrictions results in a minimum time for one operation:

$\Delta t_{min} = (Gh/c^5)^{1/2} = \sim 10^{-43}$ sec, that is Planck time $\Delta t_{cGh}$.

It means that there is the *absolute* - independent on the computer's mass - *maximum bitrate:*

$(\Delta t_{cGh})^{-1} = (c^5/Gh)^{1/2} = \sim 10^{43}$ bits per second.

The need to correct Bremermann's limit (to be consistent with GR) does not diminish the importance of his question about whether the fundamental laws of physics limit the speed of data processing. The significance of such a limitation was emphasized by Ross Ashby back in the 1960s.[4]

The quantum limits of GR and the real challenge of quantum gravity were discovered by Matvei Bronstein in his dissertation of 1935 published in two articles in 1936.[5] Analyzing quantum constraints on the measurability of gravity, he came to the conclusion that a theory of quantum gravity would require '*the rejection of a Riemannian geometry… and perhaps also the rejection of our ordinary concepts of space and time, replacing them by some much deeper and non-evident concepts*'.[6]

He had too little time to search for those 'non-evident concepts'. In 1937, at the age of 30, he was arrested and six months later executed in a Leningrad prison, as one of the millions of victims of Stalin's terror.

I am grateful to Bentzion Fleischmann for introducing me to Bremermann's question.

---

1 Bremermann, H.J. "Optimization through Evolution and Recombination," in M.C. Yovits, G.T. Jacobi and G.D. Goldstein (eds.). Self-organizing Systems. Washington, DC: Spartan Books, 1962, pp. 93-106; see: Klir, G.J. Facets of Systems Science. Springer, 2001, p. 144.

2 Gorelik, G. and V. Frenkel. Matvei Petrovich Bronstein and Soviet Theoretical Physics in the Thirties, Springer Basel AG, 2011. (Ch.5. *cGh*-physics in Bronstein's life.)

3 Gorelik, G. "First Steps of Quantum Gravity and the Planck Values," in Studies in the History of General Relativity (Einstein Studies, Vol. 3, Eds. J Eisenstaedt, A J Kox) (Boston: Birkhauser, 1992) p. 364-379.

4 Ashby, W. R. "Some Consequences of Bremermann's Limit for Information-processing Systems [1968]," Reprinted in R.C. Conant (ed.). Mechanisms of Intelligence: Ross Ashby's Writings on Cybernetics. Intersystems Publishers, 1981.

5 Bronstein, M. Republication of: Quantum theory of weak gravitational fields. Gen Relativ Gravit 44, 267–283 (2012). https://doi.org/10.1007/s10714-011-1285-4.

6 Gorelik, G. The drama of ideas in the history of quantum gravity: Niels Bohr, Lev Landau, and Matvei Bronstein // *EPJ H* **49**, 18 (2024). https://doi.org/10.1140/epjh/s13129-024-00080-9 SharedIt link: https://rdcu.be/dQ4j7.